\documentclass[epj]{webofc}

\usepackage[varg]{txfonts}   
\begin{document}
%
%
\title{gSeaGen: a GENIE-based code for neutrino telescopes}

\author{Carla Distefano\inst{1}\fnsep\thanks{\email{distefano_c@lns.infn.it}}  
        for the ANTARES and KM3NeT Collaborations}

\institute{INFN Laboratori Nazionali del Sud, Via S.Sofia 62, 95123, Catania, Italy}

\abstract{
The gSeaGen code is a GENIE based application to generate neutrino-induced events in an underwater neutrino detector. 
The gSeaGen code is able to generate events induced by all neutrino flavours, taking into account topological differences between track-type and shower-like events. The neutrino interaction is simulated taking into account the density and the composition of the media surrounding the detector. The main features of gSeaGen will be presented together with some examples of its application within ANTARES and KM3NeT.}

\maketitle

\section{Introduction}
\label{intro}

Monte Carlo simulations play an important role in the data analysis of neutrino telescopes. Simulations are used to design and optimise trigger, reconstruction and event selection algorithms. 
A reliable simulator of neutrino events at the detector is therefore required.

GENIE \cite{Andreopoulos:2009rq} is a neutrino event generator for the experimental neutrino physics community.
It has a focus on low energies and is currently used by a large number of experiments working in the neutrino oscillation field.
The goal of the project is the development of a "canonical" Monte Carlo simulating the physics of neutrino interactions whose validity extends to all nuclear targets and neutrino flavours from MeV to PeV scales. At the moment GENIE is validated up to 5 TeV \cite{GenieWeb}.

The gSeaGen code is a neutrino event generator for neutrino telescopes. It simulates particles created in all-flavour neutrino interactions, which may produce detectable Cherenkov light.
gSeaGen uses GENIE to simulate the neutrino interactions. The kinematics of the generated particles is used as input to the codes simulating the detector response \cite{annarita}. 
The code is written in C++. The main features are described in the following.

\section{The neutrino interaction volume}
\label{sec:vol} 

\begin{figure}[ht]
\centering
\includegraphics[scale=0.5]{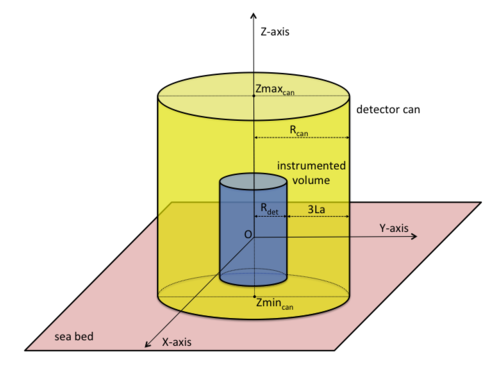}
\caption{Definition of the detector can.}
\label{fig:can}       
\end{figure}

The gSeaGen simulation code depends only on the detector size. The detector is defined inside the code by the so-called {\it can}.
It represents the detector horizon, i.e. the volume sensitive to light. Within this volume  the Cherenkov light is generated in the next steps of  
simulation to study the detector response. 
The {\it can} is a cylinder exceeding the instrumented volume by three light absorption lengths, $L_a$, bounded by the sea bed from which the light can not emerge (see Fig. \ref{fig:can}).

The {\it interaction volume} is the volume where a neutrino interaction can produce detectable particles. 
In case of electron neutrinos and neutral current interactions of muon or tau neutrinos, 
resulting particles may be detected only if they are generated inside the light sensitive volume (shower-like events) and the
interaction volume is defined as a cylinder coincident with the {\it can} and entirely made by seawater. 
If muon or tau neutrino charged current interactions are simulated, secondary muons may be detected also if the interaction vertex is outside the can (track-type events).
In this case the interaction volume is a cylinder made by a layer of rock and a layer of seawater surrounding the {\it can}. 
Its size is dimensioned according to the muon maximum range in water and in rock, evaluated at the highest energy of simulated neutrinos.  
The maximum muon range is input by the user. In the simulations reported in this work, the maximum muon range was evaluated 
through a Monte Carlo simulation using the MUSIC code \cite{Antonioli:1997qw}.

\section{The target media}
\label{sec:media}

Four different target media are defined: SeaWater, Rock (used to define the interaction volume), 
Mantle and Core (entering in the calculation of the transmission probability through the Earth). 
The composition of all media is set by the user, providing the
possibility to study the systematics due to medium compositions and also to simulate under-ice detectors.
The density profile of inner layers of the Earth (Mantle and Core media) is described according to the Preliminary Reference Earth Model (PREM) \cite{PREM}.

\section{Simulation algorithm}
\label{sec:gen}

The energy range considered for event generation is binned in equal divisions of log-energy. For each bin, 
the interaction probability is scaled up to reduce the number of trials. The probability scale is the maximum interaction probability 
(i.e. probability at maximum energy and for the maximum possible path length) summed over initial states and it is calculated by GENIE \cite{Andreopoulos:2009rq}.   

If track-type events may be generated, the interaction volume and the number of events are also scaled.   
The neutrino energy is drawn according to a power-law energy spectrum and its direction is randomly extracted according to a flat distribution in the solid angle. 
The track vertex is drawn on a circular surface (outside the interaction volume and covering its projection onto 
a plane perpendicular to each neutrino's direction). 

Once the neutrino is generated, its interaction is simulated using the GENIE event generation driver class GMCJDriver \cite{Andreopoulos:2009rq}.
The neutrino-induced particles generated inside the can are stored in the output file. Muons generated outside the can are propagated with the MUSIC code \cite{Antonioli:1997qw}
and stored if they reach the can surface.
 
\section{Calculation of the event weight}
\label{sec:wgt}

During the simulation, the code assigns a weight $W_{evt}$ to each event in order to normalise the generation to a real neutrino flux. The event weight is the product of the real neutrino spectrum 
$\phi(E_\nu,\cos\theta_\nu)$, at the generated neutrino energy $E_\nu$ and direction $\theta_\nu$, and the generation weight $W_{gen}$: 
\begin{equation}
W_{evt}=W_{gen}\cdot \phi(E_\nu,\cos\theta_\nu). \\
\end{equation}
The generation weight is defined as the inverse of the simulated neutrino spectrum:
\begin{equation}
W_{gen}= \frac{I_E \cdot I_\theta \cdot T_{gen} \cdot A_{gen} \cdot N_\nu \cdot E_\nu^X \cdot P_{scale}  \cdot P_{Earth}(E_\nu,\cos\theta_\nu)}{N_{Tot}}, 
\end{equation}
where $I_E$ and $I_\theta$ are the energy and angular phase space factors, $T_{gen}$ is the simulated time interval, $A_{gen}$ is the generation area, $N_\nu$ is the number of simulated neutrino types and $N_{Tot}$ is the total number of simulated neutrinos.
$P_{scale}$ is the GENIE interaction probability scale (see Sec. \ref{sec:gen}). 
The neutrino transmission probability through the Earth is calculated as:   
\begin{equation}
P_{Earth}(E_\nu,\cos\theta_\nu)=e^{-N_A\cdot\sigma(E_\nu)\cdot\rho_l(\theta_\nu)},
\end{equation}
where $\sigma(E_\nu)$ is the total cross section per nucleon (taking into account the different layer compositions); $\rho_l(\theta_\nu)$ is the amount of material encountered by a neutrino in its passage through the Earth. The latter is computed with the line integral 
$\rho_l(\theta_\nu)=\int_L\rho_{Earth(r)}dl$, being $L$ the neutrino path at the angle $\theta_\nu$ and $\rho_{Earth(r)}$ is the PREM Earth density profile \cite{PREM}.  

\subsection{Calculation of the systematic weights}

The accuracy of the input simulation parameters is known and the uncertainties related to nu interactions can be propagated with GENIE. 
For each input physics quantity $P$, a systematic parameter $x_P$ is introduced. Tweaking this systematic parameter modifies the corresponding physics parameter P as follows:
\begin{equation}
P\to P'= P(1 +x_P\cdot\delta P/P),
\end{equation}
where $\delta P$ is the estimated standard deviation of $P$. 
The calculation of the systematic errors in GENIE is based on an event reweighting strategy. A description of the full reweighting scheme is reported in \cite{Andreopoulos:2015wxa}. 

The evaluation of the systematics has been implemented in gSeaGen, using the GENIE class GReWeight \cite{Andreopoulos:2015wxa}. 
The implementation accepts single parameters or a list of them as input. In the latter case, the code treats all parameters at the same time and calculates the global systematic weight. 
If the calculation is activated, the systematic weights $w_{sys}$ are written in the output file. The modified distributions are obtained by multiplying the event weights by $w_{sys}$.

\section{Applications}
\label{sec:res}

The development of gSeaGen code started within ANTARES Collaboration \cite{Collaboration:2011nsa}, providing the possibility to use modern and maintained neutrino interaction codes/libraries. Currently, the code is used as a cross-check for GenHen \cite{genhen}, the standard generator code written in FORTRAN, limiting the comparison within the present GENIE validity range. gSeaGen can generate high energy events in ANTARES and KM3NeT-ARCA \cite{ARCA-ICRC2015}, when the GENIE extension at the PeV scale will be available.  At present, gSeaGen is the reference code for simulation of the KM3NeT-ORCA detector \cite{ORCA-ICRC2015}.   

\begin{figure}[ht]
\centering
\includegraphics[scale=0.55]{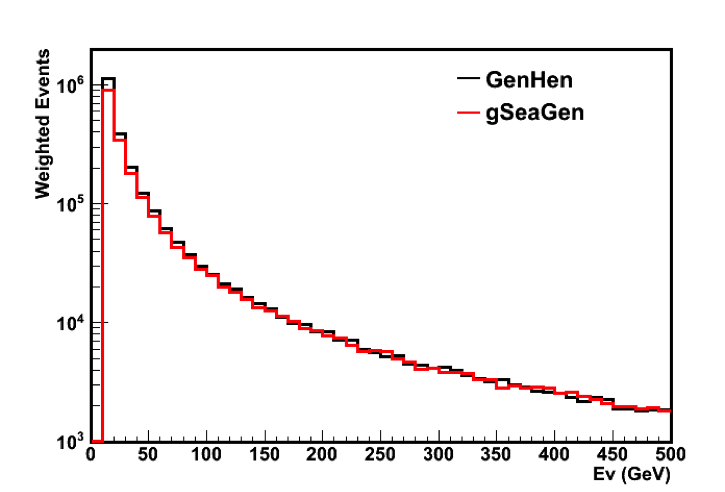}
\includegraphics[scale=0.55]{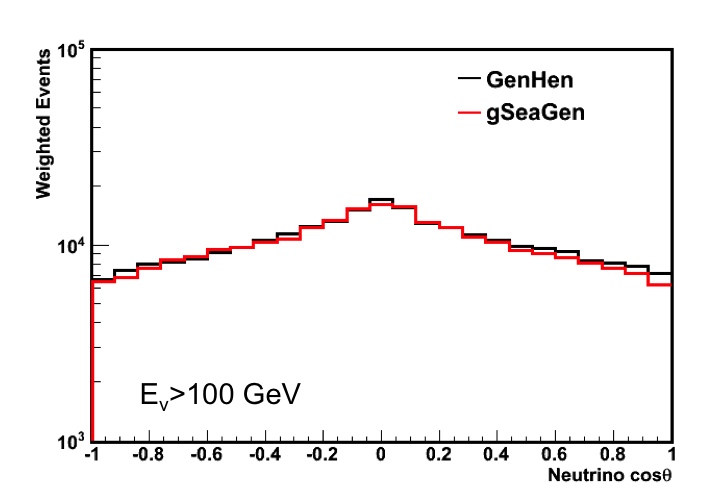}
\caption{Energy spectrum and angular distribution of atmospheric muon neutrinos and anti-neutrinos producing detectable events at the ANTARES detector can. Generated events are weighted according to the Bartol atmospheric muon flux \cite{Barr:2004br}. Results from the standard neutrino event generator GenHen are shown for comparison \cite{genhen}.}
\label{fig:muons}       
\end{figure}

As an example of application of gSeaGen, a flux of muon neutrinos and anti-neutrinos have been generated for the ANTARES detector {\it can}. The spectrum of generated events have been shaped according to the Bartol atmospheric muon flux \cite{Barr:2004br}. The results are reported in Fig. \ref{fig:muons} in terms of the energy spectrum and angular distributions of neutrinos producing detectable events (i.e. inside or reaching the detector {\it can}).

\bibliography{biblio.bib}

\end{document}